**Common high-performance semiconducting polymers are not amorphous but semi-para-crystalline**

*Sara Marina, Edgar Gutierrez-Fernandez, Junkal Gutierrez, Marco Gobbi, Eduardo Solano, Jeromy Rech, Wei You, Luis Hueso, Agnieszka Tercjak, Harald Ade and Jaime Martin\**

Sara Marina, Edgar Gutiérrez and Dr. Jaime Martín
POLYMAT, University of the Basque Country UPV/EHU Av. de Tolosa 72, 20018, Donostia-San Sebastián, Spain

Dr. Junkal Gutierrez and Dr. Agnieszka Tercjak
Group 'Materials + Technologies', Faculty of Engineering Guipuzcoa, University of the Basque Country (UPV/EHU), Plaza Europa 1, 20018 Donostia, Spain.

Dr. Junkal Gutierrez
Faculty of Engineering Vitoria-Gasteiz, University of the Basque Country (UPV/EHU), C/Nieves Cano 12, 01006 Vitoria-Gasteiz, Spain

Dr. Marco Gobbi and Prof. Luis Hueso
CIC nanoGUNE BRTA, 20018 Donostia-San Sebastian, Basque Country, Spain.
IKERBASQUE, Basque Foundation for Science, 48013 Bilbao, Basque Country, Spain.

Dr. Marco Gobbi
Centro de Física de Materiales CFM-MPC (CSIC-UPV/EHU), 20018 Donostia-San Sebastian, Basque Country, Spain

Dr. Eduardo Solano
ALBA Synchrotron Light Source, NCD-SWEET Beamline, 08290 Cerdanyola del Vallès, Spain

Jeromy J. Rech and Prof. Wei You
Department of Chemistry, University of North Carolina at Chapel Hill, Chapel Hill, North Carolina, 27599, USA

Prof. Harald Ade
Department of Physics and Organic and Carbon Electronics Laboratories (ORaCEL), North Carolina State University, Raleigh, NC, 27695 USA

Dr. Jaime Martín
Universidade da Coruña, Grupo de Polímeros, Departamento de Física e Ciencias da Terra, Centro de Investigacións Tecnolóxicas (CIT), Esteiro, 15471 Ferrol, Spain
Ikerbasque, Basque Foundation for Science, 48013 Bilbao, Spain
E-mail jaime.martin.perez@udc.es

**Precise determination of the solid-state microstructure of semiconducting polymers is of paramount importance for the further development of these materials in various organic electronic technologies. Yet, prior characterization of the ordering of semiconducting polymers often resulted in conundrums in which X-ray scattering and microscopy yielded seemingly contradicting results. Here, based on fast scanning calorimetry, we introduce for the first time the concept of the semi-para-crystallinity and measurement of the degree of para-crystallinity (ordered volume/mass fraction) in a set of materials that previously eluded understanding. In combination with lattice distortion determination within para-crystals ($g$-parameter from X-ray scattering) and nanomorphology, the complete solid-state microstructure is correlated with device properties. Our data show that the long-range charge carrier transport in these materials is more sensitive to the interconnection of para-crystal units than to the amount of structural order itself.**

1. Introduction

The ability of semiconducting polymers to transport charges has proven to be closely connected with their solid-state microstructure and, more specifically, with the presence of ordered molecular structures, such as crystals, their characteristics (i.e. size, orientation, defects, intrinsic disorder, etc.) and the interconnection between them, e.g. via tie-chains.[1]

From the solid-state microstructure standpoint, polymeric materials have been typically classified as either amorphous (glassy) or partially crystalline (i.e. semicrystalline).[2] An amorphous polymer material consist of polymer chains adopting disordered coil-like conformations resulting from the enormous number of possible rotational isomeric states of chains. In partially crystalline polymers, traditional models, e.g. the "fringed micelle model"[3] or the spherulitic model,[4] assume that amorphous regions combine with crystalline regions in which chain segments with relatively extended conformations are stacked in lamellar crystallites.[2, 5] Thus, these structural models for semicrystalline materials allow for a simple parametrization of the overall degree of structural order in terms of the degree of crystallinity, i.e. the fraction of crystalline regions to the total volume/mass of the sample.

The characteristics of the semicrystalline microstructure (e.g. semi-crystalline morphology and structure parameters) can be inferred from the combination of diffraction techniques (e.g. X-ray diffraction), microscopies (optical, electronic and force microscopies), and methods probing the melting of crystals, the most common being differential scanning calorimetry (DSC).[2] Thus, the identification and quantification of structural order in semi-crystalline polymeric materials has been traditionally realized analyzing e.g. the Bragg peaks of the X-ray diffraction patterns, measuring the birefringence under a cross-polarized optical microscope (POM), and/or quantifying the enthalpy change during the melting of crystals by DSC. Obviously, amorphous polymer materials do not feature any of the aforementioned signals.

The analysis of solution-spun lyotropic polymers, such as aramid fibers,[6] revealed, moreover, a different solid-state order – somehow between that of crystalline and amorphous materials – described by the so-called para-crystalline model.[7] Contrarily to semicrystalline materials, the overall degree of structural order in para-crystalline materials is not given by the volume fraction occupied by molecularly ordered and disordered regions, but by the quality of the para-crystalline lattice, i.e. the cumulative lattice disorder as captured in the para-crystallinity distortion parameter, $g$.

Progress in materials engineering for organic electronics has led to the development of semiconducting polymers with solid-state microstructures ranging from completely amorphous, e.g. PTAA,[1] to some with moderate-to-high degrees of structural ordering, e.g. P3HT and PBTTT.[1, 8] The degree of order in these highly-order semiconducting polymers can be readily characterized by, e.g., DSC and X-ray methods[9] like in commodity semicrystalline polymers. However, the analysis of the quality of the molecular packing in semiconducting polymers revealed that even the most ordered materials feature a fairly disordered π-stacking (e.g. *g*-parameters of 7.3% were measured for PBTTT).[10] Hence, it was proposed that ordered structural elements in many semi-conducting polymer films exhibit greater similarity with para-crystals than with regular crystals.

But unlike typical para-crystalline materials, such as aramide fibers, semiconducting polymer thin films tend to contain also significant fractions of structurally disordered regions. Hence, the *g*-parameter seems not sufficient to characterize the overall degree of structural order of these materials, as it is solely related to para-crystalline units. Indeed, this practice may have led to conundrums and misinterpretations in the past. For example, based on the *g*-parameter value, some of the current champion polymers materials for organic solar cells (OSCs), namely Poly[[4,8-bis[5-(2-ethylhexyl)-2-thienyl]benzo[1,2-b:4,5-b′]dithiophene-2,6-diyl]-2,5-thiophenediyl[5,7-bis(2-ethylhexyl)-4,8-dioxo-4H,8H-benzo[1,2- c:4,5-c′]dithiophene-1,3-diyl]] (PBDB-T), and its fluorinated (PBDB-T-2F or PM6) and chlorinated (PBDB-T-2Cl or PM7) derivatives (**Figure 1**) have been classified as structurally amorphous.[11] Moreover, the absence of optical birefringence and measurable melting signals in DSC [11] supported this conclusion. As a result, most advanced morphology-function models for OPV materials systems have been established on the premise that these polymers are structurally disordered.[12] However, distinctive aggregate-like structures are clearly revealed when these materials are inspected by electron or probe microscopies, reflecting some sort of ordered molecular packing [7a, 13] These seemingly contradictory results suggest that (*i*) *g*-parameter –alone– is not adequately reflecting the overall structural order in these polymer materials, therefore (*ii*) the actual degree of structural order –and, hence, the overall solid-state microstructure – of many of the best-performing semiconducting polymers for OPV remains still unknown.

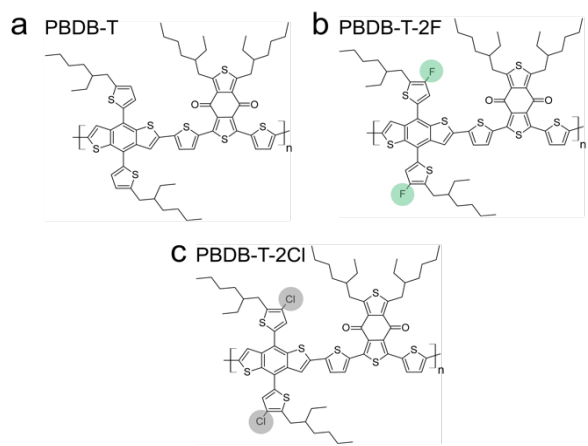

**Figure 1.** Chemical structures of (a) PBDB-T, (b) PBDB-T-2F (PM6) and (c) PBDB-T-2Cl (PM7)

In this paper, we solve this apparent paradox by introducing a concept: the degree of para-crystallinity. Analogously to the degree of semicrystallinity in semicrystalline materials, the degree of para-crystallinity features the mass/volume fraction of para-crystalline regions in semi-para-crystalline materials, without considering (in a first approximation) how ordered the molecular packing within para-crystalline units is. Hence, the characterization of the overall structural order in semi-conducting polymers for OPV would rely in two parameters: the $g$-parameter, which accounts for the lattice disorder within para-crystals, and the degree of para-crystallinity, which measures the mass/volume fraction of ordered material. (The likely existing connection between both parameters needs to be investigated in future studies).

In addition to introducing the concept, we show that the degree of para-crystallinity of device-relevant thin films can be readily obtained by fast scanning calorimetry (FSC). Hence, we demonstrate that PM6, PM7 and PBDB-T polymer films, which seem amorphous based on the $g$-parameter criteria, are in fact notably ordered, with degrees of para-crystallinity exceeding 50 % for PM7, i.e. similar to some of the most ordered semiconducting polymers like P3HT.[9a] The combination of this information with the characterization of nanomorphology and para-crystalline lattice distortion ($g$-parameter) provides complete information about the solid-state microstructure of high performing OPV materials for the first time. We thus establish that the moderate-to-high degree of para-crystallinity in PM6, PM7 and PBDB-T is underpinned by a highly dense arrangement of very small and very disordered para-crystallites. Having elucidated the semi-para-crystalline microstructure of polymer films, we investigate the interplay between the long-range charge transport and the structural features. Our data demonstrates that the field-effect charge mobility is very sensitive to the presence of small para-crystallites interconnecting primary para-crystals.

## 2. Results and Discussion

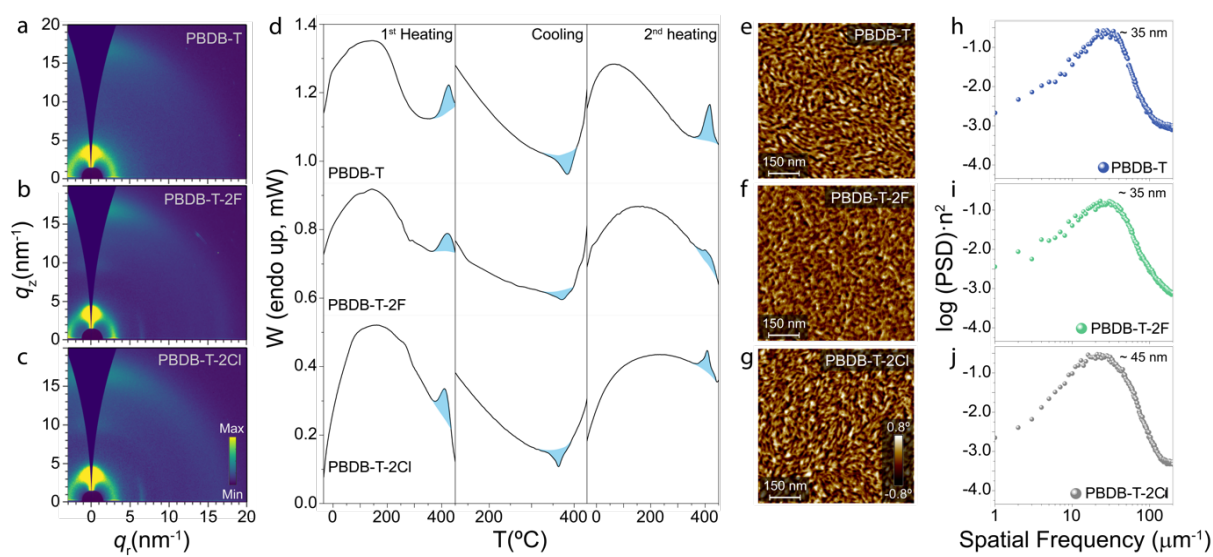

**Figure 2.** 2D-GIWAXS pattern for spin-cast (a) PBDB-T, (b) PBDB-T-2F and (c) PBDB-T-2Cl. (d) Representative fast scanning calorimetry (FSC) thermograms (raw data) for PBDB-T, PBDB-T-2F and PBDB-T-2Cl: 1st heating, cooling and 2nd heating scans are shown from left to right. Scanning rate was 4000 °C·s$^{-1}$ in all experiments. (e-g) AFM phase-contrast images and for PBDB-T, PBDB-T-2F and PBDB-T-2Cl, respectively, and corresponding power spectral density (PSD) curves (h-j).

In full accordance with previous reports,[11] when standard procedures to assess the solid-state microstructure of polymeric materials are employed, results suggest that spin cast PBDB-T, PBDBT-2F and PBDBT-2Cl films are amorphous (data shown in Figure S1, S2, S3 and Table

S1 of the Supporting Information and in Figure 2a-c). For example, films are not birefringent under POM (see Figure S1 Supporting Information); reliable endothermic peaks associated to crystal melting processes are not detected in DSC traces (supporting information Figure S2); and grazing incidence wide angle X-ray scattering (GIWAXS) patterns in **Figure 2a-c** and Figure S3 of the Supporting Information only exhibit broad diffraction maxima from the π-stacked planes, i.e. the (010), and the lamellar packing of aromatic backbones and aliphatic side chains, i.e. the (100). The analysis of the (010) and (100) diffraction peaks disclose moreover *g*-values between 16 and 21 % and between 22 and 24 % , respectively (Table S2 Supporting Information), which are well-beyond those typically found for (partially) ordered materials (which typically exhibit *g*-parameters between 0 and 12 %).[14]

However, when the spin-cast films are characterized by fast scanning calorimetry (FSC), clear endothermic signals are revealed at ~400 °C (panel labelled as "1$^{st}$ heating" in Figure 2d), i.e. at the temperatures where diffraction peaks disappear in temperature-resolved *in situ* GIWAXS experiments (Figure S4 of the Supporting Information). Hence, because we detect clear intense melting processes by calorimetry, these polymer films ought to be – at least – para-crystalline. Our data suggests thus, that GIWAXS and calorimetry disagree when traditional definitions and *g*-parameter boundary benchmarks are used. We can argue that the *g*-parameter scale based on a-SiO$_2$ reference may not be appropriate to classify semiconducting polymer thin films, and that a higher *g*-value than a 12 % needs to be established as the correct boundary benchmark for amorphous polymer semiconductors.

It should be noted that compared to regular DSC, FSC allows for the application of much faster scanning rates (scanning rates of 4000 °C·s$^{−1}$ were applied in our experiments), which on the one hand prevents thermal degradation of compounds at high temperatures, thereby enabling the detection of order-disorder transitions, and, on the other hand, allow to probe spin cast films that can be safely correlated with GIWAXS data (and device data).

The subsequent FSC cooling sweeps for the three polymer films show exothermic peaks due to the formation of para-crystals upon cooling. This highlights (a) that para-crystals can be also formed by cooling the molten semiconducting polymers and (b) that the formation of para-crystals is an extremely efficient and fast process (in this experiment para-crystals are formed during cooling at -4000 °C·s$^{-1}$). Interestingly, moreover, the structural characteristics of the para-crystals developed by cooling the melt must be quite similar to those of solution-processed spin cast films, because 1$^{st}$ and 2$^{nd}$ heating scans exhibited endothermic processes alike in enthalpy and position.

Because the enthalpy change during the melting of spin cast para-crystals can be measured from the FSC 1$^{st}$ heating scan, we can quantitatively estimate the mass fraction of material that has suffered the order-disorder transition, i.e. the fraction of para-crystalline material. By analogy with semicrystalline materials, we name this parameter the degree of para-crystallinity. The degree of para-crystallinity can be obtained by normalizing the measured enthalpy to that of the 100% para-crystalline P3HT[31] (in the absence of specific values for the polymer materials analyzed here). More details of this calculation are provided in Supporting Information Figures S5-S7. The degree of para-crystallinity thus obtained amount to ~38% for PBDB-T, ~27% PBDB-T-2F and ~54% PBDB-T-2Cl which, despite of the uncertainty of the methodology, suggests that the PBDB-T family of polymers exhibit a remarkable molecular order, indeed, similar to that of P3HT films[9a] (note: the value for PBDB-T-2F is most probably underestimated as the maximum temperature covered in the experiment, i.e. 450 °C, is not high enough to resolve completely the endothermic peak and moreover).

Measured values for the degree of crystallinity imply the existence of significant fractions of disordered material together with the para-crystals. Hence, unlike previously reported para-crystalline materials like aramide fibers, which do not include discrete amorphous regions and thus the degree of structural disorder can be parametrized solely by the *g*-parameter, the degree of disorder in the polymers analyzed here must have two contributions: a first contribution associated with the distortion on the para-crystalline lattice (like the one in aramide fibers) and a further contribution associated with a non-para-crystalline, disordered material fraction. In other words, the microstructure of these semiconducting polymers is not adequately represented by the para-crystalline model, which must be adjusted to include non-para-crystalline regions. Hence, in order to describe the microstructure of these polymer materials we introduce the concepts of semi-para-crystallinity and semi-para-crystalline microstructure.

The question that must be answered in the first place is how this semi-para-crystalline microstructure is; and more specifically, how is it possible that such a remarkable fraction of ordered material is detected by FSC, while materials are not birefringent and exhibit such high $g$-parameter values. In order to shed light on this matter we investigated the morphology of the thin films in the length-scale of a few tens of nm by grazing incidence small angle X-ray scattering (GISAXS) and atomic force microscopy (AFM). The GISAXS data (included in Figure S8 of the Supporting Information) and, more clearly, the AFM phase-contrast images displayed in Figure 2e-g, suggest a multiphasic nanomorphology that is compatible with the coexistence of small para-crystalline regions separated by more disordered regions. Both power spectral density (PSD) analysis of the AFM images (shown in Figure 2h-j) and GISAXS analysis (Figure S8 of the Supporting Information) indicate moreover that the solid-state microstructure in these polymers has an average characteristic size of 35-45 nm, which is most likely related to the spacing between para-crystallites, while para-crystallites are 20-30 nm. (The complete AFM study is included in the Figure S9-S11 of the Supporting Information).

Hence, the semi-para-crystalline microstructure relies on the presence of a very dense arrangement of small para-crystallites. Within these para-crystallites, molecular packing lacks of long-range order, or much order at all along both the lamellar direction and the π-stacking direction, as evidenced by GIWAXS and discussed in the Figure S12 of the Supporting

Information. A schematic depicting this solid-state microstructure is shown in Figure S13 of the Supporting Information and in a later section of the manuscript in **Figure 5**.

Because of the high-performance of these polymers in OSCs, an important question arises as to whether the semi-para-crystalline microstructure is retained in donor:acceptor blends of OSC devices. If so, currently accepted morphology-function and degradation models will need to be adjusted to include a measurable semi-para-crystallinity. In order to investigate this point, we analyzed a PBDB-T:ITIC (1:1) binary blend produced using similar processing conditions as those employed in efficient device preparation (i.e. spin casting a chlorobenzene solution containing 0.5% 1,8-diiodooctane, DIO, followed by a thermal annealing at 160 °C).[15] Like neat polymer films, photovoltaic blends seem to be amorphous when they are characterized by GIWAXS (**Figure 3b** and Figure S14). However, an endothermic feature showing up at about 400 °C in the FSC heating scan (shadowed in blue in Figure 3a) reflects here again the melting of PBDB-T para-crystals, revealing para-crystalline order also in donor domains of binary blends. Moreover, the multiphasic morphology that is characteristic of the semi-para-crystalline microstructure of these polymers is also revealed in the AFM phase image (Figure 3c and Figure S15). Note: the endothermic peak at about 260 °C is due to calorimetric signals from ITIC domains.[16]

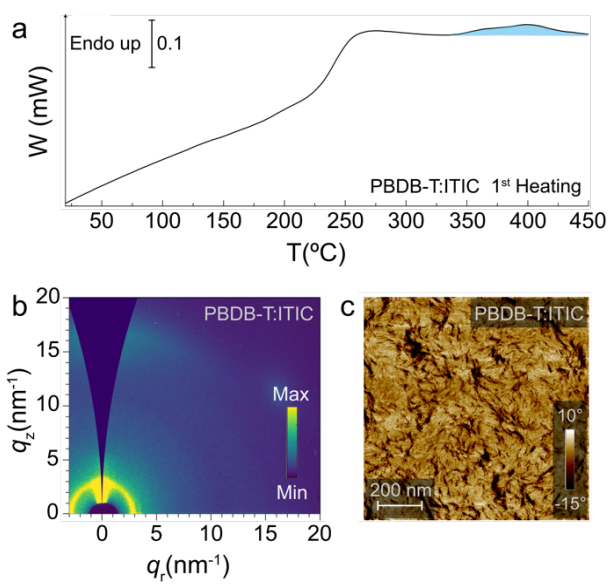

**Figure 3.** PBDB-T:ITIC (1:1) films processed as in devices:[22] (a) FSC first heating thermogram (4000 °C·s$^{-1}$), (b) *Ex situ* 2D-GIWAXS pattern and (c) AFM phase image.

Motivated by our results, we also investigated whether our finding can be extended to other semiconducting polymers with efficient photoconversion capability that, based primarily on X-ray diffraction data, have been so far considered poorly-ordered or amorphous, such as Poly [[4,8-bis[(2-ethylhexyl)oxy]benzo[1,2-b:4,5-b']dithiophene-2,6-diyl][3-fluoro-2-[(2-ethylhexyl)carbonyl]thieno[3,4-b]thiophenediyl ]] (PTB7) and Fluorobenzotriazole-Based Polymers (FTAZ).[17] Once again, FSC results included in (Supporting Information Figure S16), provide compelling evidence that PTB7 and FTAZ thin films are semi-para-crystalline both right after being solution processed and when cooled down from the melt. Hence, we speculate that the solid-state microstructure described in this paper might be a common feature among thin films of many high-performing rigid polymers employed for OPVs, although the detailed interplay between thermal and diffraction characterization has to be more fully explored and understood.

Having decoded the microstructure of spin-cast PBDB-T, PBDB-T-2F and PBDB-T-2Cl films, we set on to investigate the interplay between the characteristic aspects of the semi-para-crystalline microstructure and the charge transport. For that, organic field-effect transistors (OFETs) were fabricated with semi-para-crystalline films exhibiting microstructural modifications resulting from thermally annealing the films at different temperatures ($T_a$s). The selected $T_a$s were 25 °C, 160 °C and 280 °C and the resulting solid-state microstructures, referred to as *A*-, *B*-, and *C*-microstructures, respectively, are displayed in **Figure 4**. (We recall that *A*-microstructure is also achieved when in spin-cast films and films cooled down from the melt at 4000 °C·s$^{-1}$).

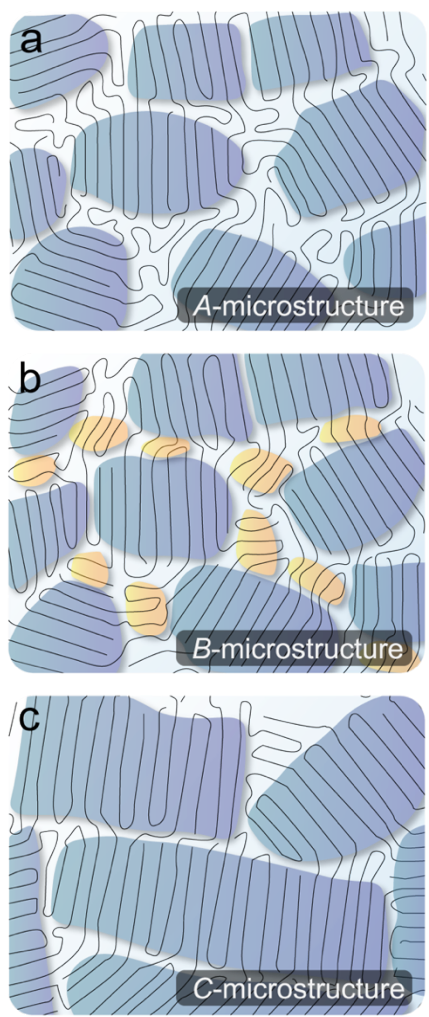

**Figure 4**. Schematics depicting the semi-para-crystalline microstructure of as-cast films (and films annealed at 25 ºC, denoted as *A*-microstructure) (a) and films annealed at 160 ºC (b, denoted as *B*-microstructure) and 280 ºC (c, denoted as *C*-microstructure)

**Figure 5a**, 5b and 5c show representative transfer characteristics for the PBDB-T, PBDB-T-2F and PBDB-T-2Cl transistors, respectively, and their field-effect mobility values, $\mu$, are summarized in Figure 5d (the entire set of device data is included in Figure S17 of the Supporting Information). While a 2.5-to-4-fold increase in $\mu$ is found for transistors exhibiting *B*-microstructure compared to those with *A*-microstructure, the comparison between transistors having *B*- and *C*-microstructures shows uneven results: maximum $\mu$-values are recorded for PBDB-T-2F and PBDB-T-2Cl transistors with *C*-microstructures, but in the case of PBDB-T transistors, the optimized $\mu$-values are found for the *B*-microstructure.

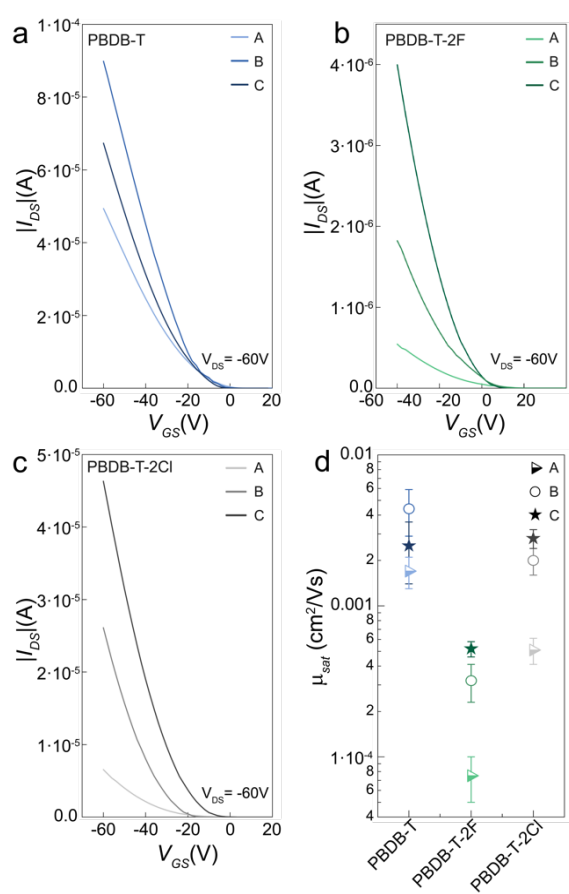

**Figure 5.** Representative transfer characteristics for (a) PBDB-T, (b) PBDB-T-2F and (c) PBDB-T-2Cl field-effect transistors annealed at 25 ºC, 160 ºC and 280 ºC. (d) Field-effect mobility, µ, values for the different polymer samples. All transfer characteristics were measured in transistors with width W = 10000 and length L = 40.

The higher $\mu$-values for *B*-microstructures compared to *A*-microstructures in all polymers results from a more efficient long-range charge transport due to the presence of small para-crystals between primary para-crystals (in *B*-microstructure, see Figure 4b), which enhance the percolation of the structural network that sustain the charge transport. Some indications that support this view are obtained from the FSC, AFM and GIWAXS data. **Figure 6a**, 6b and 6c show the FSC heating traces for PBDB-T, PBDB-T-2F and PBDB-T-2Cl thin films exhibiting *B*-microstructures (thick black lines) and *A*-microstructures (thin grey lines). Because no changes are observed in the melting peaks of preexisting para-crystals, we rule out significant secondary (para-)crystallization and recrystallization processes in the polymer films during thermal annealing at 160 °C. Interestingly, curves for *B*-microstructures exhibit additional weak endothermic peaks between 200 and 300 °C that we associate with the melting of new para-crystallites grown during the annealing[18] (we note that the kinetics of these fresh para-crystal formation is discussed in Figure S18 and S19 of the Supporting Information). The new para-crystallites must be even smaller and/or more defective than the preexisting ones because they exhibit melting temperatures that are more than 100 °C lower. Resulting in a relatively minor increase of the overall degree of para-crystallinity (~10%), the new para-crystallites induce a notable enhancement of the charge transport of semi-para-crystalline ($\mu$ increases between 2.5 and 4 times depending on the polymer film). Hence, we argue that the new para-crystallites may grow within non-ordered regions between preexisting para-crystals, thereby improving their interconnection pathways. This idea is in line with the charge transport model by Noriega, Rivnay and Salleo (Figure 5b),[1] and furthermore highlights that long tie-chains are not the only possible mechanism of interconnection of ordered regions, which can occur also through small para-crystallites.

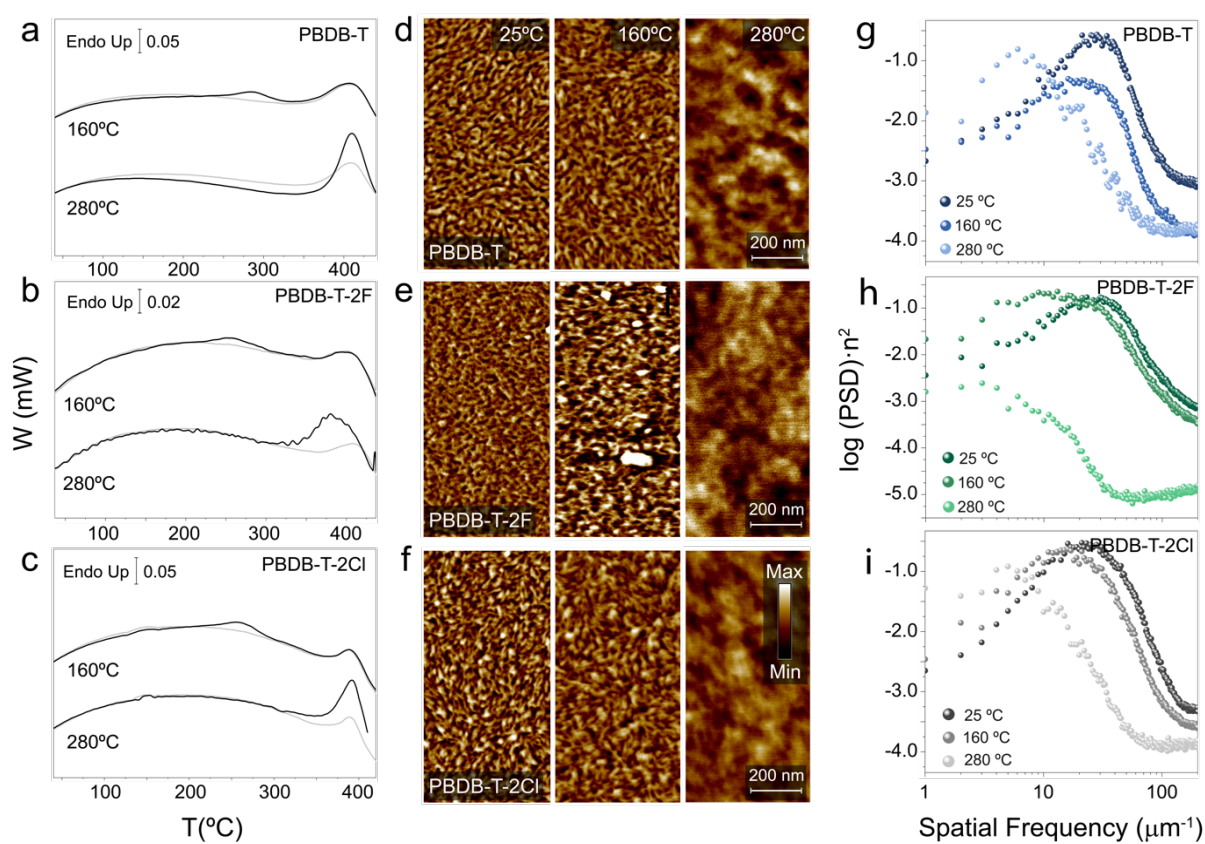

**Figure 6.** Fast scanning calorimetry thermograms (raw data) of (a) PBDB-T, (b) PBDB-T-2F and (c) PBDB-T-2Cl: thick black lines correspond to heating scans for films annealing for 2h at 160ºC (*B*-microstructure) and 280 ºC (*C*-microstructure). Heating curves for *A*-microstructures (obtained by cooling down form the melt at 4000 ºC·s$^{-1}$) are shown as grey lines. Atomic force microscopy phase images for (d) PBDB-T *A*-, *B*- and *C*-microstructures, (e) PBDB-T-2F *A*-, *B*- and *C*-microstructures and (f) PBDB-T-2Cl *A*-, *B*- and *C*-microstructures. (g-i) Power spectral density (PSD) curves of AFM phase contrast mages.

Apart from a minor increase of the characteristic length-scale of the microstructure suggested by the PSD analysis (Figure 6g-6i), the general nanomorphology of *A*- and *B*-microstructures is pretty similar when inspected by AFM (Figure 6d-6f and Figures S20-S22 of the Supporting Information). In the same line, GIWAXS experiments (shown in **Figure 7** and Figures S23-S25 of the Supporting Information) revealed a minor reduction of *g* for the (100) and (010) planes in *B*-microstructures (summarized in Table S2 of the Supporting Information and Figure 7d and 7e) [we note that the (010) planes are expected to be connected with the overlap of π-orbitals and thereby linked to preferential electronic pathways]. *C*-microstructures featured, however, substantial morphological and structural changes compared to *A*- and *B*-microstructures. FSC curves for C-microstructure show, for example, larger enthalpy changes during para-crystals melting, thus larger degrees of para-crystallinities, which reach 47 % for PBDB-T and 62 % for PM7. (The value for PM6 cannot be safe extracted from our calorimetric data). This increase of the degree of para-crystallinity coincides with a significant enlargement of the para-crystals, which reach >100 nm, as probed by PSD analysis (Figure 6g-6i) and AFM images (Figure 6d-6f and Figures S20-S22 of the Supporting Information). Furthermore, the *g*-values for the (100) and (010) planes are significantly lower in *C*-microstructures, reflecting a rapid reduction of the lattice disorder when films are annealed at high temperatures (*g*-values are summarized in Figure 7d and 7e and Table S2 of the Supporting Information). We also noted that the *g*-values for the (002) planes are fairly insensitive to the temperature in the temperature range analyzed.

An increase of the degree of para-crystallinity and the size of para-crystals aligned with a reduction of lattice distortion can explain the enhancement of $\mu$ observed for PBDB-T-2F and PBDB-T-2Cl (Figure 4d). However, the opposite trend is found for PBDB-T. Interestingly, our texture analysis in Figure 7g ($A_{ip}/A_{oop}$ refers to the ratio between the (100) diffraction maxima along the out-of-plane and the in-plane directions) highlights that PBDB-T para-crystals suffer from a progressive change of orientation as the $T_a$ increases, yielding face-on oriented crystallites at the higher $T_a$s analyzed, i.e. those leading to *C*-microstructures. Face-on oriented para-crystals have their (010) planes parallel to the substrate, which may hinder the in-plane charge transport, resulting in the low mobility observed for the PBDB-T *C*-microstructure.

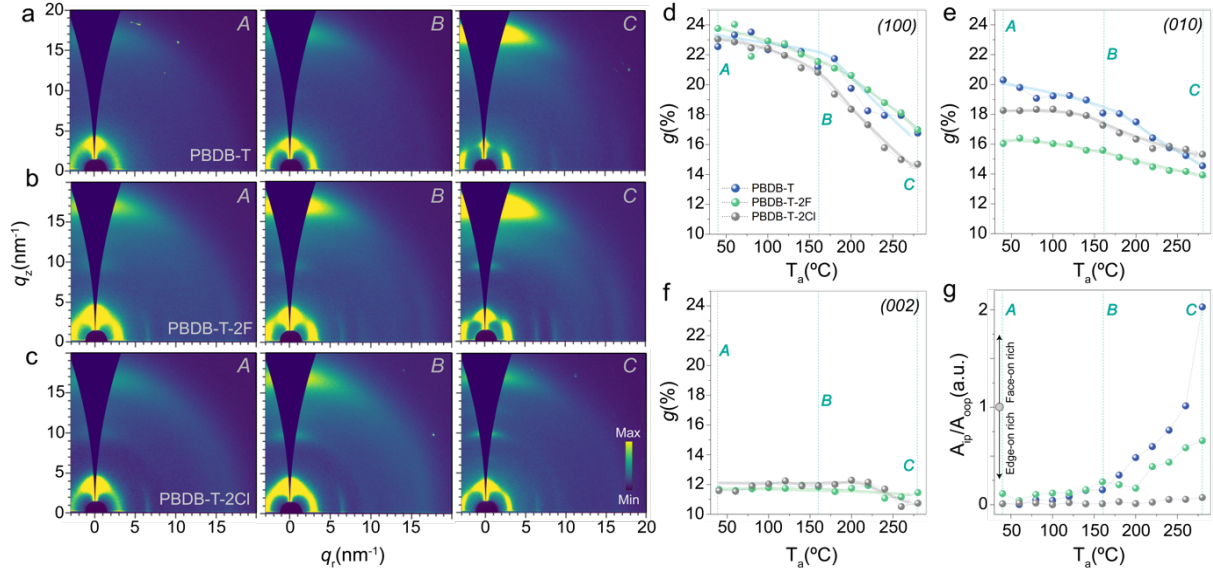

**Figure 7.** *Ex situ* 2D-GIWAXS pattern of (a) PBDB-T *A*-, *B*- and *C*-microstructures; (b) PBDB-T-2F *A*-, *B*- and *C*-microstructures; and (c) PBDB-T-2Cl *A*-, *B*- and *C*-microstructures. Paracrystalline distortion parameter, *g*, for the (d) (100) planes, (e) (010) planes and (f) (00*l*) planes as a function of the annealing temperature, $T_a$. (g) Ratio between the (100) diffraction maxima along the out-of-plane and in-plane directions ($A_{ip}/A_{oop}$), for the different polymers as a function of $T_a$.

## 3. Conclusions

In this paper, we shed light on the long-standing debate about structural order in high-performing semiconducting polymers. So far, the characterization of the amount of structural order in these materials resulted in conundrums, in which scattering and microscopy methods led to seemingly contradicting conclusions. Thus, based on most-standardized X-ray analysis these materials have been assumed to be amorphous and thus have been considered in most advanced morphology-function and stability prediction models for OSCs. Employing fast scanning calorimetry, we discover and demonstrate, however, that they can be in fact notably ordered. The misinterpretation originates from employing the *g*-parameter as the only criteria to assess structural order of semiconducting polymers. These materials often exhibit *g*-parameter values that are similar to those of amorphous a-SiO$_2$, which can lead to mistakenly interpret that these materials are amorphous. Though para-crystalline lattices are indeed largely distorted in these materials (hence the large *g*-parameter values), we find a surprisingly small fraction of total polymeric material that is actually disordered. Thus, we conclude that the precise assessment of structural order in semiconducting polymers requires information about -at least- two parameters: the *g*-parameter, which accounts for the para-crystalline lattice distortion, and the degree of para-crystallinity (introduced here for the first time), which measures the mass/volume fraction of para-crystalline material from the analysis of order-disorder peaks in FSC scans. The relationship between both parameters needs to be analyzed in future studies.

We demonstrate that the remarkably high degree of structural order in these semi-para-crystalline polymers is underpinned by a dense arrangement of very small and disordered para-crystallites that coexist with more disordered regions. The disordered fractions of semi-para-crystalline materials are reasonably expected to exhibit a glass transition process. This too needs to be researched and our work opens up an avenue of quantitative and much more complete understanding of complex materials that semi-conducting polymers represent. Partially in line with the charge transport model of semicrystalline polymers by Noriega, Rivnay and Salleo[1], the field-effect charge mobility of semi-para-crystalline polymer materials is very sensitive to the interconnection of para-crystal units through further small para-crystallites. The reduction of the para-crystalline lattice disorder, the degree of para-crystallinity and the para-crystallite´s orientation are also proven to impact charge transport.

Because the semi-para-crystalline microstructure is likely to be a common feature among many polymers having semi-rigid backbones and some amphiphilicity, including some kinds of biomacromolecules, we believe that our results can have direct implications not only in the organic electronics arena, where e.g. device operation models must now include the degree of para-crystallinity, but also in other now seemingly unconnected fields.

4. **Experimental Methods**

*Materials*: The polymer donors PBDB-T, PBDB-T-2F, PBDB-T-2Cl and PTB7 were supplied by Ossila Ltd. FTAZ was synthesized as previously reported. [43] Chlorobenzene (CB) and chloroform (CHCl$_3$) were purchased from Merck and used as received. Unless indicated, PBDB-T and PBDB-T-2Cl were spin cast from 20 mg·mL$^{-1}$ chlorobenzene solutions at 2000 rpm during 60s and PBDB-T-2F was spin cast from 16 mg·mL$^{-1}$ chloroform solutions at 3000 rpm.

*Fast Scanning Calorimetry (FSC):* FSC measurements were carried out in a Mettler Toledo Flash DSC 1 equipped with a two-stage intracooler, allowing for temperature control between -90 and 450 °C and nitrogen purge (75 mL·min$^{-1}$ N$_2$ gas flow). MultiSTAR UFS1 (24 × 24 × 0.6 mm$^3$) MEMS chip sensors were conditioned and corrected according the specifications prior to use. For the experiments, polymer solutions were spin cast onto the backside of the chip sensor. For analysis of the 1$^{st}$ heating scan of spin coated films, samples were scanned between -90 °C and 450 °C at 4000 °C·s$^{-1}$. 2$^{nd}$ heating scans were obtained in the same experimental conditions after the film were cooled down from 450°C to -90 °C at 4000 °C·s$^{-1}$

*Grazing Incidence Wide Angle X-Ray Scattering (GIWAXS):* GIWAXS measurements were performed at the BL11 NCD-SWEET beamline at ALBA Synchrotron Radiation Facility (Spain). The incident X-ray beam energy was set to 12.4 eV using a channel cut Si (1 1 1) monochromator. The angle of incidence $\alpha_i$ was set between 0.1-0.15° to ensure surface sensitivity. The scattering patterns were recorded using a Rayonix® LX255-HS area detector, which consists of a pixel array of 1920 × 5760 pixels (H × V) with a pixel size of 44 × 44 μm$^2$. Data are expressed as a function of the scattering vector ($Q$), which was calibrated using Cr$_2$O$_3$ as standard sample, obtaining a sample to detector distance of 200.93 mm. Temperature-resolved *in situ* X-ray experiments were performed using a *Linkam® THMS 600* stage adapted for grazing incidence experiments. The heating rate used was 20 °C·min$^{-1}$ and the temperature difference between frames was 4 °C. Sample alignment was automatically performed each 50 °C. Exposure times for *in situ* and *ex situ* experiments were 1 and 5 s, respectively. All the measurements were performed under N$_2$ atmosphere to minimize the damage of the films. 2D GIWAXS patterns were corrected as a function of the components of the scattering vector. Samples for GIWAXS were spin cast on Si wafers. The thermal treatments were performed in a *Linkam* hot stage under N$_2$ atmosphere. Edges of the samples were removed to eliminate edge effects in the GIWAXS pattern.

*Grazing incidence X-ray scattering (GISAXS):* The GISAXS experiments were conducted at NCD-SWEET beamline of ALBA synchrotron (Spain). A monochromatic X-ray beam with an energy of 12.4 keV was shone on the samples with an incidence angle of 0.12° and 0.15°. The exposure time for room temperature measurements was 1 s and the sample to detector distance was 2.54 m. The 2D patterns were recorded with a Pilatus3 S 1 M detector, which consists of a pixel array 1043 × 981 (V × H) pixels of 172 × 172 μm$^2$. Horizontal line $q_y$ cut profiles were done at the Yoneda peak. For GISAXS experiments samples were spin cast on Si wafers. The thermal treatments were performed in a *Linkam* hot stage under N$_2$ atmosphere. Edges of the samples were removed to eliminate edge effects in the GISAXS scattering pattern.

*Atomic force microscopy (AFM).* AFM images were obtained using a scanning probe microscope (Dimension ICON, Bruker) under ambient conditions. Tapping mode was employed in air using an integrated tip/cantilever (125 μm in length with ca. 300 kHz resonant frequency). Scan rates ranged from 0.7 to 1.2 Hz s$^{-1}$. Measurements were performed with 512 scan lines and target amplitude around 0.9 V. The amplitude setpoint for all investigated samples was ~300 mV. Different regions of the samples were scanned to ensure that the morphology of the investigated materials is the representative one. In the AFM, phase images, the brighter regions correspond to areas of the sample with higher modulus (hard domains) and darker regions correspond to softer areas. Samples were spin cast from 20 mg·mL$^{-1}$ solution at 2000 rpm during 60s. The thermal treatments were performed in a *Linkam* hot stage under N$_2$ atmosphere. Samples were annealed for 10 minutes at selected temperature and quenched at 50ºC·min$^{-1}$ to room temperature.

Quantitative information about the para-crystallite size was obtained from power spectral density (PSD) analysis applied on the AFM phase-contrast images using the NanoScope Analyses 1.9 software.

***Field-effect transistor fabrication and measurement.*** Field-effect transistors in a bottom gate, bottom contact geometry were fabricated by spin coating the different polymers onto Si/SiO$_2$ (150 nm) substrates with prepatterned pairs of Ti/Au (5 nm/ 36 nm) electrodes. The n++ Si substrate was used as gate electrode, the SiO$_2$ (150nm) as gate dielectric and the Ti/Au contacts as source (S) and drain (D). The prepatterned pairs of electrodes with interdigitated geometry were fabricated using conventional lithographic techniques. Channel length (L) and widths (W) varied in the range L ~ 5-50 μm and W ~ 5000-10000 μm.

The individual chips were cleaned with acetone and later with isopropanol. Before spin coating the polymers the chip were introduced in the ozone cleaner (supplied by by Ossila Ltd.) for 1h. Samples were spin cast from 10 mg·mL$^{-1}$ solution at 2000 rpm during 60s. The thermal treatments were performed in a *Linkam* hot stage under N*2* atmosphere. Samples were annealed for 10 minutes at selected temperature and quenched at 50ºC·min$^{-1}$ to room temperature.

The electrical characteristics of the transistors were measured using a Keithley 4200-SCS semiconductor analyzer connected to a variable temperature Lakeshore probe station. The measurements were carried out with the samples in high vacuum, and the samples were left 12 hours in vacuum prior to the measurements, to minimize the effect of oxygen doping.

To extract the field-effect mobility, the transfer curves were analyzed using standard field-effect transistor equations for the saturation regime:

$$I_{DS} = -\frac{W}{2L}\mu C(V_{GS} - V_{Th})^2 \quad (1)$$

Where $I_{DS}$ is the drains source current, μ is the mobility, C is the capacitance per unit area, $V_{GS}$ is the gate-source voltage and $V_{Th}$ is the threshold voltage. In each chip, more than 10 devices were tested to extract the average mobility, and the error is given by the standard deviation.

**Supporting Information**

Supporting Information is available from the Wiley Online Library or from the author.

**Acknowledgements**

S. Marina is grateful to POLYMAT for her PhD scholarship. J.M. thanks MICINN /FEDER for the Ramón y Cajal contract and the grant Ref. PGC2018-094620-A-I00). The Basque Country Government is also acknowledged for the grant Ref. PIBA19-0051. J.M and E.F.-G. acknowledge support through the European Union's Horizon 2020 research and innovation program, H2020-FETOPEN-01-2018-2020 (FET-Open Challenging Current Thinking), 'LION-HEARTED', grant agreement n. 828984. J.M would like to thank the financial support provided by the IONBIKE RISE project. This project has received funding from the European Union's Horizon 2020 research and innovation programme under the Marie Skłodowska-Curie grant agreement No 823989. J.J.R and W.Y acknowledge the financial support of the National Science Foundation (CBET-1934374). The authors thank Marek Grzelczak and Joscha Kruse for their support with UV-Vis spectroscopy measurements. The authors also thank the technical and human support provided by SGIker of UPV/EHU and the European funding (ERDF and ESF). The work at CIC nanoGUNE was supported by "la Caixa" Foundation (ID 100010434), under the agreement LCF/BQ/PI19/11690017 and by the Spanish MICINN under the Maria de

Maeztu Units of Excellence Programme (MDM-2016-0618) and Project RTI2018-094861-B-100 and PID2019-108153GA-I00.

Received: ((will be filled in by the editorial staff))
Revised: ((will be filled in by the editorial staff))
Published online: ((will be filled in by the editorial staff))**ToC figure**

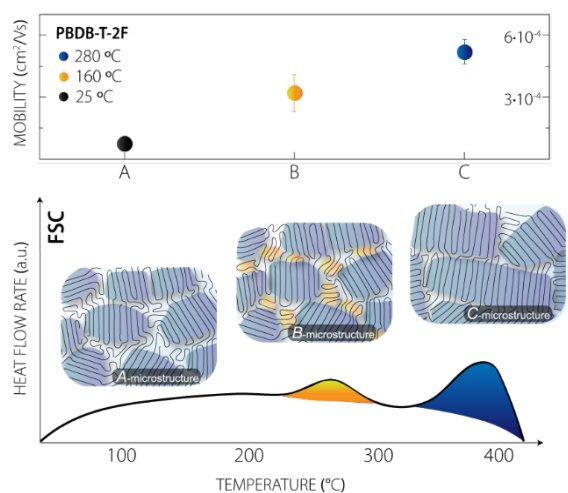

**References**

[1]   R. Noriega, J. Rivnay, K. Vandewal, F. P. V. Koch, N. Stingelin, P. Smith, M. F. Toney, A. Salleo, *Nature Materials* **2013**, *12*, 1038.
[2]   G. Strobl, *The Physics of Polymers: conceps for understanding their structures and behaviour.*, Springer, Berlin Heidelberg New York **2007**.
[3]   Hermann K, G. O, A. W, *Z physik Chem B* **1930**, *10*.
[4]   F. Van Antwerpen, D. W. Van Krevelen, *Journal of Polymer Science: Polymer Physics Edition* **1972**, *10*, 2409.
[5]   J. Martín, A. Iturrospe, A. Cavallaro, A. Arbe, N. Stingelin, T. A. Ezquerra, C. Mijangos, A. Nogales, *Chemistry of Materials* **2017**, *29*, 3515.23